\documentclass[galaxies,article,submit,moreauthors,pdftex]{Definitions/mdpi}

\def\prd{Phys. Rev. D}
\def\prl{Phys. Rev. Lett.}
\def\jcap{JCAP}
\def\jhep{J. High Energy Phys.}

\usepackage{color,xcolor,graphicx,amsmath,amssymb,mathtools,hyperref,cleveref,subcaption,booktabs,multirow,geometry,ulem, physics, slashed, ulem, cancel}
\newcommand{\lt}{\left }
\newcommand{\rt}{\right }
\newcommand{\stb}{ \bf \color{blue!30!red} }

\newcommand{\be}{\begin{eqnarray}}
\newcommand{\ee}{\end{eqnarray}}
\newcommand{\baln}{\begin{aligned}}
\newcommand{\ealn}{\end{aligned}}

\newcommand{\bal}{\begin{align}}
\newcommand{\eal}{\end{align}}

\definecolor{shilaviolet}{rgb}{0.8, 0.0, 0.8}
\newcommand{\Porey}{\textcolor{shilaviolet}}

\firstpage{1} 
\pubvolume{1}
\issuenum{1}
\articlenumber{0}
\pubyear{2021}
\copyrightyear{2020}
\datereceived{} 
\dateaccepted{} 
\datepublished{} 
\hreflink{https://doi.org/} 
\Title{Effects of 2HDM in electroweak phase transition}

\TitleCitation{}

\Author{Arnab~Chaudhuri $^{1,\ddagger,}$*, Maxim~Yu.~Khlopov $^{2,3,4,\ddagger}$\orcidM{} and Shiladitya~Porey $^{5,}$\orcidS{}}

\address{%
$^{1}$ \quad Department of Physics, Novosibirsk State University,
Pirogova 2, 630090, Novosibirsk, Russia; arnabchaudhuri.7@gmail.com\\
$^{2}$ \quad CNRS, Astroparticule et Cosmologie, Université de Paris, F-75013 Paris, France2; khlopov@apc.in2p3.fr\\
$^{3}$ \quad Institute of Physics, Southern Federal University, 344090 Rostov on Don, Russia\\
$^{4}$ \quad Center for Cosmopartilce Physics “Cosmion”, National Research Nuclear University “MEPhI” (Moscow Engineering Physics Institute), 115409 Moscow, Russia\\
$^{5}$ \quad Department of Physics, Novosibirsk State University, Pirogova 2, 630090, Novosibirsk, Russia; shiladitya@g.nsu.ru\\
}

\corres{Correspondence:  arnabchaudhuri.7@gmail.com
}

\secondnote{These authors contributed equally to this work.}

\abstract{
The entropy production scenarios due to electroweak phase transition (EWPT) in the framework of the minimal extension of standard model namely two Higgs doublet model(2HDM) is revisited. The possibility of first order phase transition is discussed. Intense parameter scanning is done with the help of BSMPT, a C++ package. Numerical calculations  are performed in order to calculate the entropy production with numerous benchmark points.}


\begin{document}
\nolinenumbers

\section{Introduction}

Recently, muon g-2 experiment at FERMILAB~\cite{Abi:2021gix} measured the magnetic moment of muon and the result has drawn the attention because the value of the magnetic moment seems to be  incongruous/anomalous as it deviates from the theoretical value predicted in the Standard Model (SM). At tree level QED, the gyromagnetic ratio of muon is $g_\mu=2$. Adding virtual particles and loop corrections, the calculated value of $\lt(g_\mu -2\rt)/2$ in SM is $116591810(43)  \times 10^{-11}$ while the measured value from FERMILAB is $116592061(41) \times 10^{-11}$\cite{Arcadi:2021yyr,Chen:2021jok}. The detection of  this anomaly actually has buttressed the previous claims about the disaccord between the theoretical and experimental results~\cite{Tanabashi:2018oca}. This enduring disagreement may be the indication 
of the existence of some theories beyond the standard model (BSM) e.g. supersymmetric theory~\cite{Endo:2021zal}.


On the other side, Electroweak Phase Transition (EWPT) in the early hot universe in the SM framework with single Higgs field is also a well established theory. According to this theory, 
when the temperature of the universe drops near to a critical value ($T_c$, for details see ref.~\cite{Katz:2014bha} or sec.\ref{sec.2}), another minimum of the Higgs potential for non zero value of Higgs field appears and the phase transition ($SU(2)_L \times U(1)_Y \rightarrow U(1)_{em}$) happens and consequently the intermediate bosons and fermions gain masses.  The parameter-space of this EWPT might be altered 
if EWPT happens in
the BSM model. Here we consider the 2HDM model which provides the minimal phenomenological description of some effects of the supersymmetric model predicting two Higgs boson doublet.
For more motivation of this theory see~\cite{Barroso:2013zxa, Karmakar}.
Some extensions of 2HDM can not only produce dark matter (DM) but have the potential to successfully explain the $g_\mu$ anomaly of muon~\cite{CarcamoHernandez:2021iat}.
In this work, we discuss one of the thermodynamical properties, namely entropy production during EWPT by scanning the parameter space in details for 2HDM scenarios.

At very primeval time when the temperature of universe $T\lt(t\rt)\gg T_c$, the universe was dominated by ultra-relativistic species and $\Gamma/{\cal H}\lt(t \rt)\sim T\lt( t\rt)^{-1}\ll 1$ where $\Gamma$ is the rate of particle interaction with each others and and with the photons, ${\cal H}(t)$ is the Hubble parameter. All the species, therefore, were in thermal equilibrium with photons. During that aeon, almost all of the fermions and bosons were massless and contribution from those who were already massive (specifically speaking, decoupled components e.g. decoupled DM particles like Axion) to the total energy density of the universe was insignificant. During that time, $T\gg \lt(\mu_{\rm chm. \, pot.} -m \rt)$,  $\mu_{\rm chm.\, pot.}$ is the chemical potential of a particle
~\cite{Boeckel:2007tz} and thus massless bosons had zero chemical potential and assuming chemical potential of the fermions to be negligible 
%
the entropy density per comoving volume $s\lt( t\rt)$ was conserved and is expressed as

\begin{eqnarray}
\label{eq:entropy-conservation}
s\lt( t\rt)\equiv\frac{\rho_r\lt( t\rt) +P_r \lt( t\rt)}{T\lt( t\rt)}a\lt(t \rt)^3 = {\rm const.}
\end{eqnarray}
where $\rho_r\lt( t\rt)$ and $P_r \lt( t\rt)$ are the energy and pressure density of the
relativistic components respectively and $a\lt(t \rt)$ is the scale factor at that moment. For relativistic particles
\begin{eqnarray}
\label{eq:rho+P}
s\lt(t \rt)\sim g_{\star,s} \, T\lt(t \rt)^3
\end{eqnarray}
$g_{\star,s}$ is the  effective number of degrees of freedom in entropy and it is not constant over time. Although, it depends on the components of the primordial hot soup, it is not always same as $g_\star$ which is effective number of relativistic degrees of freedom (for discussions, see~\cite{Lunardini:2019zob}). However, for our this project, we have assumed $g_{\star,s}\approx g_\star$. Now, from eq.~\eqref{eq:entropy-conservation} and Eq.~\eqref{eq:rho+P} we get $T \sim a^{-1}$. With the expansion and cooling down of the universe, at some epoch, the universe went to the state of thermal inequilibrium and hence the value of $s$ and thus $g_\star \left(T\right) a^3 T^3$, might have increased as entropy can only either increase or remain constant.



During EWPT, as the temperature drops below from $T_c$ upto the  decoupled temperature of a particular component of the relativistic plasma, that component decouples and becomes nonrelativistic and massive. The decoupled temperature depends on the masses of the components and their respective coupling constants. This decoupling process causes the change in $g_{\star}$ of the relativistic plasma. Electroweak baryogenesis can happen during EWPT and hence thermal inequilibrium is established in the universe following Sakharov's principle~\cite{AS}. And this results in the violation of entropy conservation law and a net influx of entropy is generated. 
This led to increase of $s$. The main contribution to this influx comes from the single heaviest particle with $m(T)< T$. Hence for these temperature range, $g_\star (T)T=\text{const}.$ and the entropy rises by $a^3T^3$. Since the final temperature $T=m(T_f)$, below which a new particle starts to dominate is not dependent on $g_{\star}$, we can say the entropy increased in the background of $g_{\star}a^3 T^3$.


This paper is arranged as follows: In section \ref{sec.2}, we give the theoretical framework of EWPT in 2HDM. In section \ref{num} detailed about BSMPT package and numerical calculations as well as the results of this work are presented and it is followed by a generic conclusion in section \ref{sec.Conclusion}.

\section{Theoretical framework: EWPT theory in 2HDM}
\label{sec.2}
The Lagrangian density of EWPT theory within 2HDM is:

\begin{equation}
\mathcal{L}= \mathcal{L}_f  + \mathcal{L}_{\rm Yuk} + \mathcal{L}_{\rm gauge, kin}  + \mathcal{L}_{\rm Higgs, \, kin} - V_{\rm tot}(\Phi_1,\Phi_2, T) \label{eq: Total lagrangian}
\end{equation}

The first four terms on the right hand side of eq.~\eqref{eq: Total lagrangian} are

\begin{align} 
&\mathcal{L}_f =\sum_{j}i \lt(\bar{\Psi}^{(j)}_L \slashed{D}\Psi^{(j)}_L + \bar{\Psi}^{(j)}_R \slashed{D}\Psi^{(j)}_R \rt) \\ 
&\mathcal{L}_{\rm Yuk} = -\left[ y_e \bar{e_R} \Phi_a^\dagger L_L + y_e^* \bar{L_L} \Phi_a^\dagger e_R  + \cdots \right]  \label{eq:Yukawa lagrangian}\\
& \mathcal{L}_{\rm Higgs} = (D^\mu \Phi_a)^\dagger (D_\mu \Phi_a)\\
&\mathcal{L}_{\rm gauge, kin} = -\frac{1}{4}G^j_{\mu \nu}{G^j}^{\mu \nu}-\frac{1}{4}F^B_{\mu \nu}{F^B}^{\mu \nu}  \label{eq:gauge kin}
%
\end{align}

where $i=\sqrt{-1}$, $\Psi^{(j)}$ is the Dirac field for the $j$-th fermion species and subscript $L$ ($R$) for the left (right) chiral field and subscpit $a=1,2$ for two Higgs field respectively.  
In eq.\eqref{eq:Yukawa lagrangian}, $y_e$ is the Yukawa coupling and the sum is over all fermions, not only electron and position as shown in that equation. $G^i_{\mu \nu}$, $F^B_{\mu \nu}$ and the operators $\slashed{D}$ and $D$ are defined as
\begin{align}
    & G^j_{\mu \nu}=\partial_\mu W^j_\nu-\partial_\nu W^j_\mu - g \epsilon^{jkl}W_\mu^k W_\nu^l     \label{eq:Gmunu}\\
    & F^B_{\mu \nu}=\partial_\mu B_\nu-\partial_\nu B_\mu \\
    & \slashed{D} \Psi^{(j)}_{L,R} \equiv \gamma^\mu (\partial_\mu + ig W_\mu + i g'Y_{L,R} B_\mu) \Psi^{(j)}_{L,R}\\
    & D \equiv (\partial_\mu  + ig T^j W^j_\mu + i g'Y B_\mu )   \label{eq:D operator}
\end{align}

where Greek indices in the subscript run from $0$ to $3$. Superscript $j,k,l$ in eqns \eqref{eq:gauge kin},\eqref{eq:Gmunu} and \eqref{eq:D operator} can take values from $1$ to $3$. $g$, $g'$ are coupling constants, $\gamma^\mu$ are the gamma matrices, $W_\mu$ and $B_\mu$ are two gauge bosons.  $T^i$ is the generator of $SU(2)_L$, is also a form of Pauli matrices and $Y$ is the hypercharge generator of the $U(1)$.


The CP-conserving real type-I 2HDM potential of eq.\eqref{eq: Total lagrangian} can be expressed as:
\begin{equation} \label{2HDM potential}
    V(\Phi_1,\Phi_2,T)=V_{tree}(\Phi_1,\Phi_2)+V_{CW}(\Phi_1,\Phi_2)+V_{T}(T),
\end{equation}
where the terms respectively are the tree level potential, the Coleman-Weinberg and the temperature corrections respectively. The individual terms are described in the following equations.

The tree level potential is given by:
\begingroup\makeatletter\def\f@size{10}\check@mathfonts
\def\maketag@@@#1{\hbox{\m@th\normalsize\normalfont#1}}%
\begin{align} 
V_{\rm tree}(\Phi_1,\Phi_2)   =& m_{11}^2 \Phi_1^\dagger \Phi_1 + m_{22}^2 \Phi_2^\dagger \Phi_2 - \left[m_{12}^2 \Phi_1^\dagger \Phi_2 + m_{12}^* \Phi_2^\dagger \Phi_1 \right] + \frac{1}{2} \lambda_1 \left(\Phi_1^\dagger \Phi_1\right)^2 \nonumber \\ 
&+  \frac{1}{2} \lambda_2 \left(\Phi_2^\dagger \Phi_2\right)^2  + \lambda_3 \left(\Phi_1^\dagger \Phi_1\right)\left(\Phi_2^\dagger \Phi_2\right) + \lambda_4 \left(\Phi_1^\dagger \Phi_2\right)\left(\Phi_2^\dagger \Phi_1\right) \nonumber \\
&+ \left[\frac{1}{2} \lambda_5  \left(\Phi_1^\dagger \Phi_2\right)^2 + \frac{1}{2} \lambda_5^*  \left(\Phi_2^\dagger \Phi_1\right)^2 \right].
\end{align}
\endgroup
Here the $\lambda$s are the quartic coupling constants and all of them (including $\lambda_5$) are assumed real for this work  and are given by:

\begin{align}
    \lambda_1 &= \frac{1}{v_{\rm sm}^2 \cos^2 \beta}  \lt( - \mu^2 \tan\beta + m_h^2 \sin^2 \alpha + m_H^2 \cos^2\alpha\rt) \\
    \lambda_2 &= \frac{1}{v_{\rm sm}^2 \sin^2 \beta} \lt( - \frac{\mu^2}{\tan\beta} + m_h^2 \cos^2 \alpha + m_H^2 \sin^2\alpha\rt)   \\
    \lambda_3 &= \frac{1}{v_{\rm sm}^2 }  \lt( - \frac{2 \mu^2}{\sin 2\beta} + 2 m_{H_\pm}^2 + \lt(m_H^2 - m_h^2 \rt)\frac{\sin 2 \alpha}{\sin 2 \beta}\rt) \\
    \lambda_4 &= \frac{1}{v_{\rm sm}^2 }  \lt(  \frac{2 \mu^2}{\sin 2\beta} + m_A^2 - 2 m_{H_\pm}^2 \rt) \\
    \lambda_5 &= \frac{1}{v_{\rm sm}^2 }  \lt( \frac{2 \mu^2}{\sin 2\beta} - m_A^2  \rt)
\end{align}

The mass of $m_h, m_{H_\pm}, m_H, m_A$ and also of the other particles depends on the location of second minimum i.e. on $(v_1,v_2)$. 
Here $\alpha$ and $\beta$ are the mixing angles and $v_{\rm sm}$ is the vacuum expectation value in the SM.
The details about scanning of the parameter space  can be found in the recent works~\cite{Eberhardt:2020dat,Karmakar:2020mds}.

The other correction terms of the potential are defined respectively as follows:
\begin{align}
&V_{\rm CW}\left(v_1+v_2 \right)=\sum_j \frac{n_j}{64\pi^2} (-1)^{2s_j}m_j^4\left(v_1,v_2\right)\left[ \log\left( \frac{m_j^2 \left(v_1,v_2 \right)}{\mu^2} \right) - c_j \right] \\
&V_T=\frac{T^4}{2\pi^2}\left( \sum_{j={\rm bosons}} n_j J_{B}\left[\frac{m_j^2(v_1,v_2)}{T^2}\right]  + \sum_{j={\rm fermions}} n_j J_{F}\left[\frac{m_j^2(v_1,v_2)}{T^2}    \right] \right)
\end{align}

where $\mu$ is the  renormalisation scale which we take to be 246~GeV. As mentioned earlier, all the masses $m_j\lt( v_1, v_2\rt)$, $n_j$ and $s_j$ are mentioned in Appendix B of~\cite{Chaudhuri:2021agl}.


$J_B$ and $J_F$ are approximated {\stb in} Landau gauge up to leading orders:
\begin{eqnarray}
T^4 J_B \left[\frac{m^2}{T} \right] &&= -\frac{\pi^4 T^4}{45} + \frac{\pi^2}{12}T^2 m^2 - \frac{\pi}{6}T (m^2)^{3/2} - \frac{1}{32}m^4 \ln \frac{m^2}{a_b T^2} + \cdots,   \\
T^4 J_F \left[\frac{m^2}{T} \right] &&=  \frac{7\pi^4 T^4}{360} - \frac{\pi^2}{24}T^2 m^2 - \frac{1}{32}m^4 \ln \frac{m^2}{a_f T^2} + \cdots, 
\end{eqnarray}
where $a_b=16a_f=16\pi^2 \exp(3/2-2\gamma_E)$ with $\gamma_E$ being the Euler--Mascheroni constant.

As the temperature drops down below the critical temperature $T_c$, a second local minimum 
at $(\Phi_1=v_1,\Phi_2=v_2)$ 
appears with the same height of the global minimum situated at $\expval{\Phi_1}=\expval{\Phi_2}=0$ and $T_c$ is determined by 
 \begin{equation}
     V_{\rm tot}\left(\Phi_1=0,\Phi_2=0,T_c \right)= V_{\rm tot}\left(\Phi_1=v_1,\Phi_2=v_2,T_c \right).
 \end{equation}

It is assumed that dark matter and other components might have been present but they did not contribute much to the energy density of the universe during the particular epoch of EWPT which happened during radiation domination.

The early universe was flat, hence the metric $g_{\mu \nu}=(+,-,-,-)$.  
The energy density of the homogeneous classical field $\Phi$: 

\begin{align}
	\rho =&\partial^0 \Phi_a^\dagger \partial^0\Phi_a - (\mathcal{W}^0 \Phi_a)^\dagger \mathcal{W}^0\Phi_a 
	- (\mathcal{W}^j \Phi_a)^\dagger \mathcal{W}_j\Phi_a  \nonumber \\
	&+\lt[V_{\rm tot}(\Phi_1,\Phi_2, T)  - \mathcal{L}_{\rm gauge, kin} - \mathcal{L}_f  - \mathcal{L}_{\rm Yuk}  
	\rt]     \label{Eq: Expression for energy desnity}
\end{align}

where due to the condition of homogeneity and isotropy, all the spatial derivatives of Higgs fields are zero. Similarly,

\begin{eqnarray}
\rho + P&&= 2\partial^0\Phi_a \partial^0 \Phi_a^\dagger - i ({\mathcal{W}^0}\Phi_a)^\dagger (\partial_0 \Phi_a) + i (\partial^0{\Phi_a}^\dagger)\mathcal{W}_0\Phi_a
\label{Faltu equation 5}
\end{eqnarray}

The Higgs fields start to oscillate around the second minimum, the minimum that appeared during the EWPT. Particle production from this oscillating field causes the damping. The characteristic time of decay is equal to the decay width of the Higgs bosons. If it is large in compared to 
the expansion and thus the universe cooling rate, then we may assume that Higgs bosons essentially live in the minimum of the potential. This has being clearly discussed in~\cite{Chaudhuri:2017icn}.

Following the above assumption to be valid,
\begin{eqnarray} \label{22}
\rho &=& \Dot{\Phi}_{a, {\rm min}}^2
+  V_{\rm tot}(\Phi_1,\Phi_2, T) + \frac{g_* \pi^2}{30} T^4.
\end{eqnarray}

The 
 last term in Equation (\ref{22}) arises from the Yukawa interaction between fermions and Higgs bosons and from
the energy density of the fermions, the gauge bosons and the interaction between the Higgs and gauge bosons. This is the energy density of the relativistic particles which have not gained mass till the moment of EWPT. Essentially, the energy density of the plasma consists of two parts, namely, the energy density of the fields at the minima and the relativistic matter sector.

To calculate the entropy production, it is necessary to solve the evolution equation for energy density conservation, 
\begin{equation} \label{fried}
\Dot{\rho}=-3\mathcal{H}(\rho+P).
\end{equation}

From henceforward computational analysis was used for further calculations which are discussed in the next section.\\

\section{Numerical calculations} \label{num}
In this section we give the details about the numerical calculations which were conducted. In passing by we discuss about the C++ programming package that we used for the calculations.

BSMPT~\cite{Blinov:2015vma} is C++ tool used for calculating the strength of the electroweak phase transition in extended Higgs sectors. This is based on the loop-corrected effective potential, also including daisy re-summation of the bosonic masses. The program calculates the vacuum expectation values of the potential(VEV) ($v$) as a function of temperature and in particular, it's value at the critical temperature $T_c$. The models implemented within this tool includes CP-conserving 2HDM and next-to-minimal 2HDM. In this work we restrict ourselves to the real sector of CP-conserving 2HDM potential. 

\begin{figure}[H]
  \centering
  \begin{minipage}[b]{0.3\textwidth}
    \includegraphics[width=\textwidth]{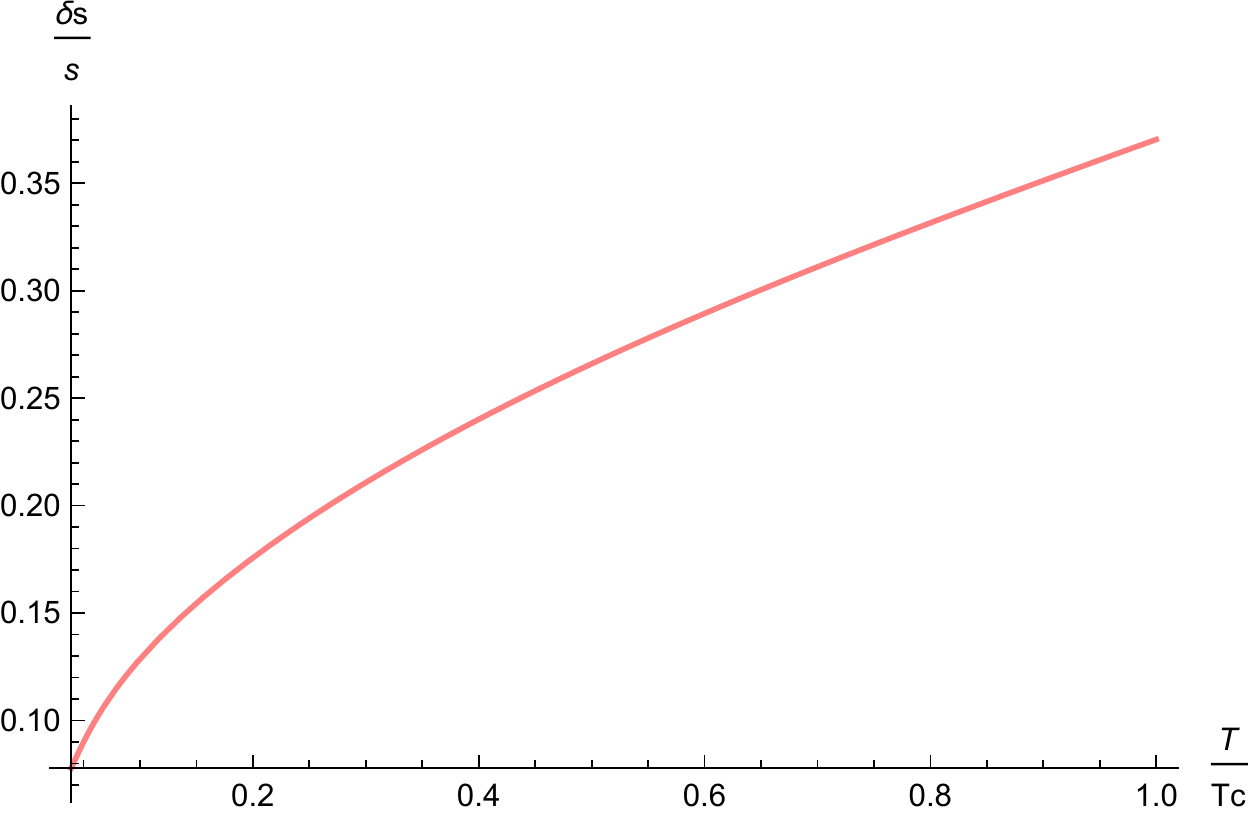}
  \end{minipage}
  \hspace*{.1cm}
  \begin{minipage}[b]{0.3\textwidth}
    \includegraphics[width=\textwidth]{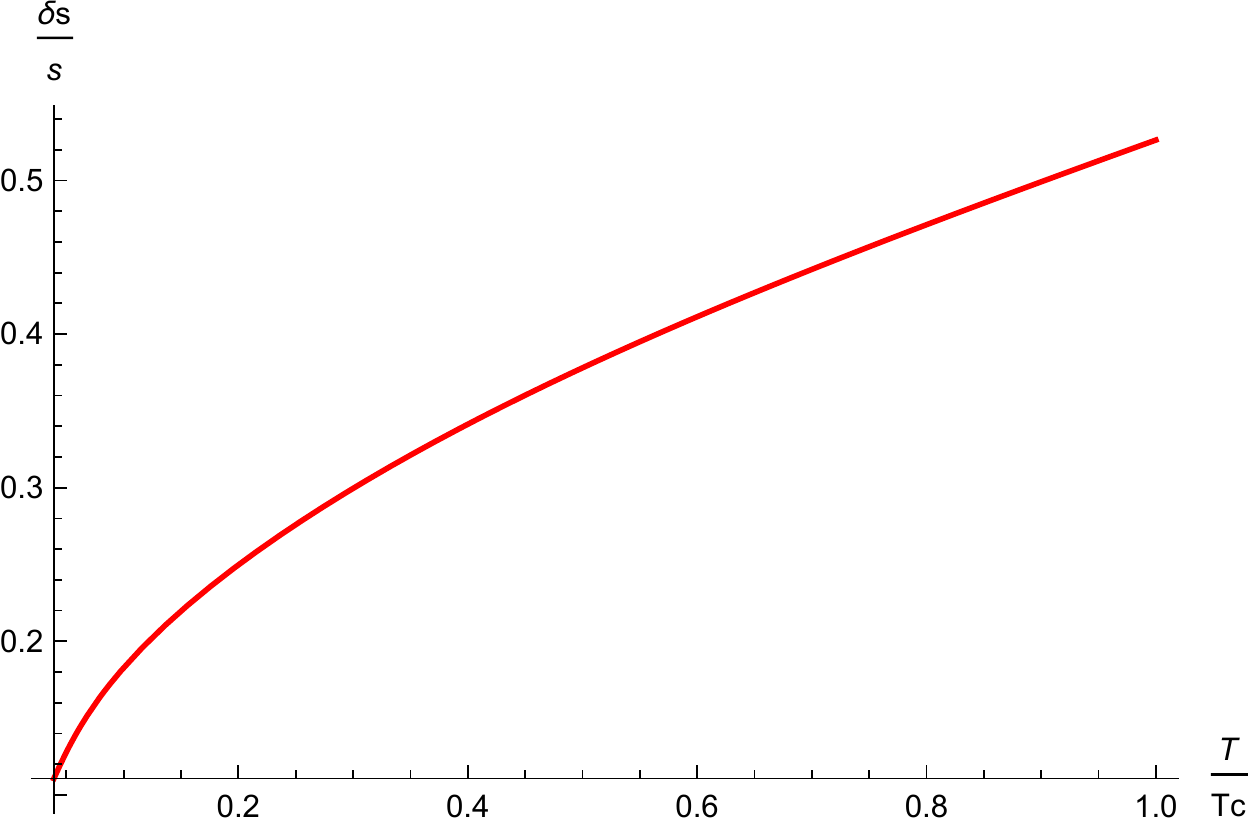}
      \end{minipage}
    \hspace*{.1cm}
    \begin{minipage}[b]{0.3\textwidth}
    \includegraphics[width=\textwidth]{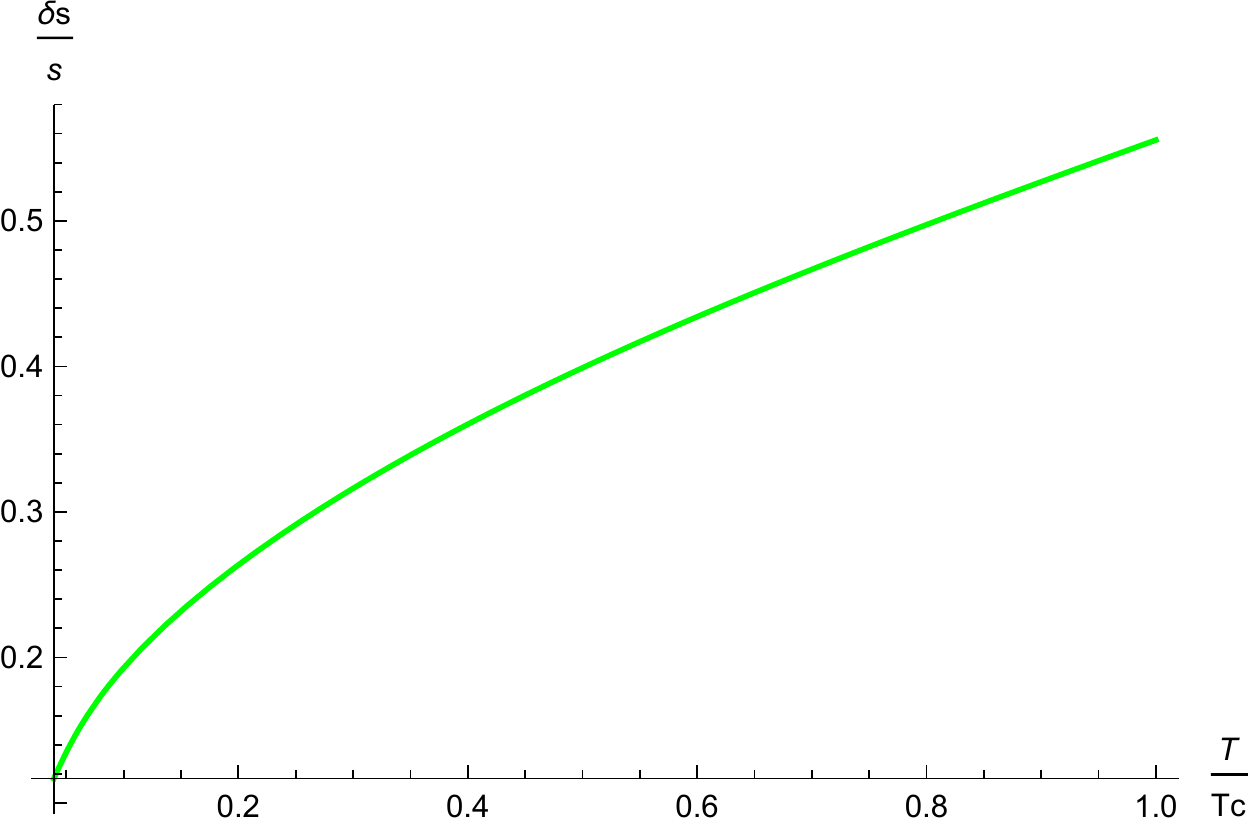}
      \end{minipage}
      \begin{minipage}[b]{0.3\textwidth}
    \includegraphics[width=\textwidth]{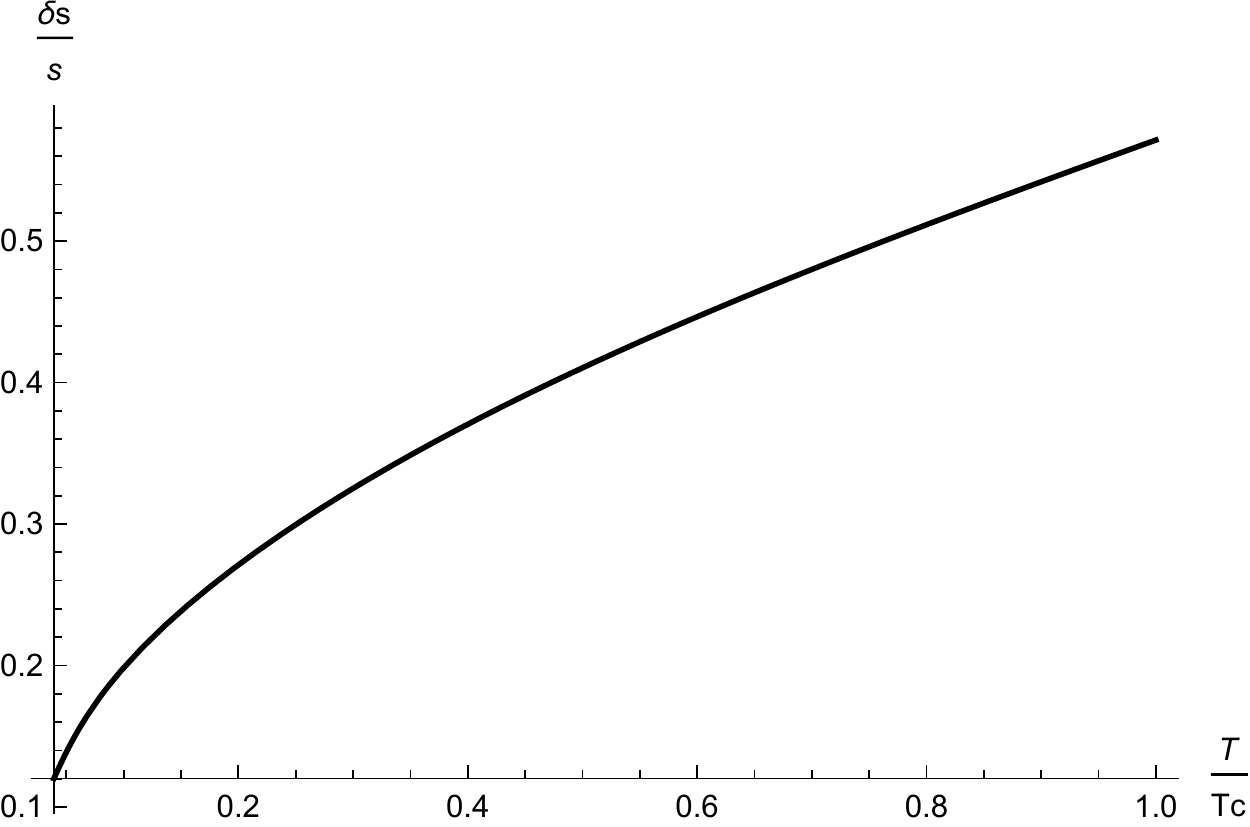}
      \end{minipage}
      \begin{minipage}[b]{0.3\textwidth}
    \includegraphics[width=\textwidth]{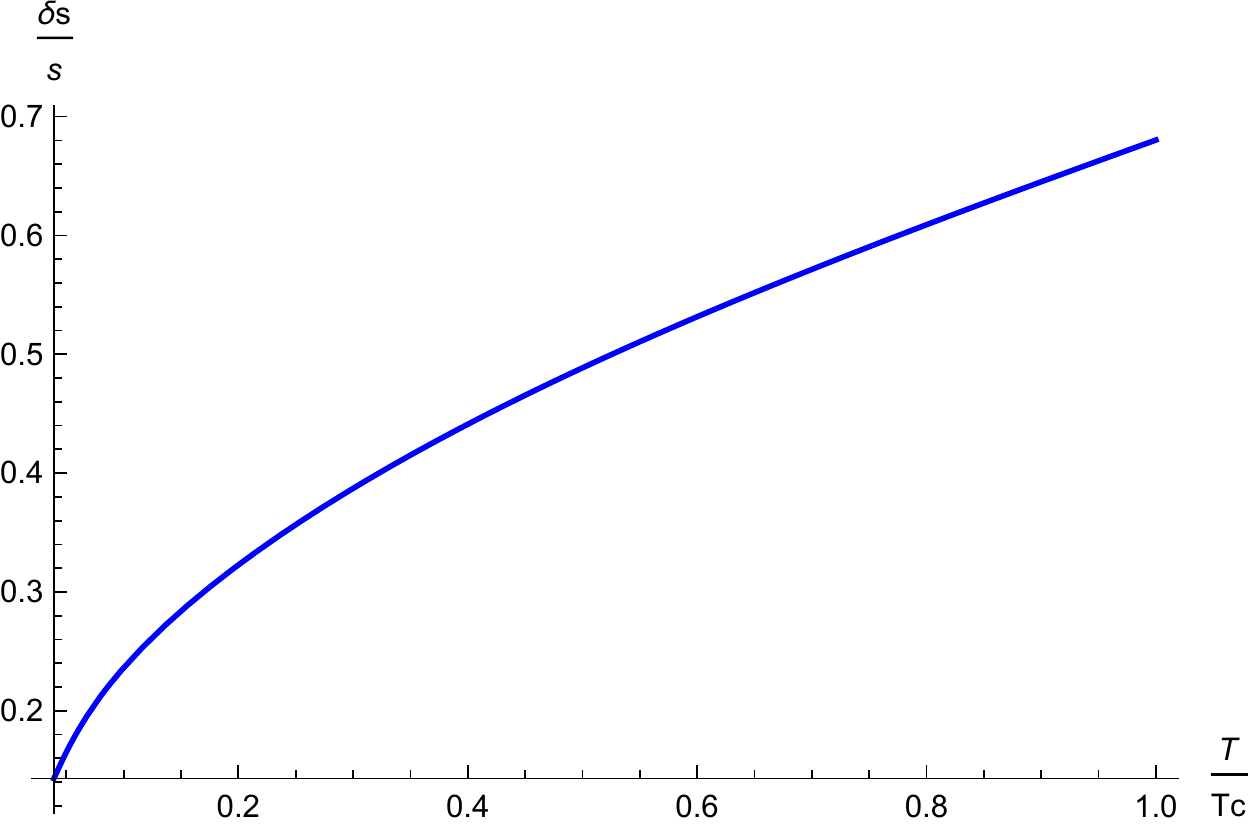}
      \end{minipage}
  \caption{The figures show the entropy production for five different benchmark(BM) points: Pink line (BM 23, $T_c=135.38~GeV$ and $\delta s/s=37\%$), Red line (BM 10,  $T_c=153.27~GeV$ and $\delta s/s=53\%$), Green line (BM 1,  $T_c=161.36~GeV$ and $\delta s/s=30\%$), Black line (BM 15,  $T_c=168.61~GeV$ and $\delta s/s=59\%$) and Blue line (BM 20  $T_c=230.18~GeV$ and $\delta s/s=70\%$).}
 \label{f-entropy-1}
 \end{figure}

This tool is used to calculate the VEVs, $T_c$s and the effective value of potential $V_{eff}(T)$s for each benchmark points in Table \ref{Table:1 Benchmark values}. The parameters are chosen in such a way that they satisfy the limiting conditions for type-I real 2HDM. It should also be noted that we have assumed that $VEV/Tc > 0.02$. The differential equation in eq.\eqref{fried} is solved numerically for all the benchmark points in Table \ref{Table:1 Benchmark values} and the entropy production is calculated for each of them as shown in Table~\ref{Table:1 Benchmark values}. Some selected figures showing this production are given in figure ~\ref{f-entropy-1}.

\begin{table}[h!] 
	\centering
	\begin{center}
		\caption{2HDM Benchmark points for entropy production}\label{Table:1 Benchmark values}
		\resizebox{0.7\textwidth}{!}{\begin{tabular}{ |c|c|c|c|c|c|c|c|c|c|c|c|c|c|c|} 
			\hline
			& $m_h$[GeV]& $m_H[GeV]$ & $m_{H^{\pm}}$[GeV]& $m_A$[GeV] & $\tan\beta$ & $\cos \left(\beta-\alpha \right)$ & $m_{12}^2~GeV^2$ & $\lambda_1$ & $\lambda_2$ & $\lambda_3$ & $\lambda_4$ & $\lambda_5$ & $T_c$ & $\delta s/s[\%]$\\
			\hline
BM1 & $125$ & $500$ & $500$ & $500$ & $2$ & $0$ & $10^5$ & $0.258$ & $0.258$ & $0.258$ & $0$ & $0$ & $ 161.36$ & $57$ \\
\hline
BM2 &" &" &" & " &" & $0.06$ & " &$1.14$ & $0.037$ & $0.63$ & $0$ & $0$ & $ 167.95$ & $59$ \\
			\hline
BM3 & " & " & " & " & $10$ & $0$ & $24752.5$ & $0.258$ & $0.258$ & $0.258$ &$0$ & $0$ &  $161.02$ & $56$\\
 \hline
 BM4 & "& " & " & " & $10$ & $0.1$ & $24752.5$ & $4.13$ & $0.22$ & $4.15$ & $0$ & $0$ & $255.71$ & $73$\\
 \hline
BM5 & " & " & " & $485$ & $2$ & $0.0$ & $10^5$ & $0.26$ & $0.26$ & $0.26$ & $-0.244$ & $0.244$ & $161.53 $  & $57$\\
 \hline
BM6 & " & " & " & $485$ & $2$ & $0.07$ & $10^5$ & $1.28$ & $0.002$ & $0.7$ & $-0.244$ & $0.244$ & $169.81 $  & $60$\\
 \hline
BM7 & " & " & " & $477$ & $2$ & $0.07$ & $10^5$ & $1.28$ & $0.002$ & $0.7$ & $-0.37$ & $0.37$ &  $169.53 $ & $60$ \\
 \hline
BM8 & " & " & " & $485$ &  $10$ & $0.0$ & $24752.5$ & $0.258$ & $0.258$ & $0.258$ & $-0.244$ & $0.244$ & $160.76 $ & $56$ \\
 \hline
 BM9 & " & " & " & $350$ & $10$ & $0.1$ & $24752.5$ & $4.13$ & $0.22$ & $4.15$ & $-2.1$ & $2.1$ & $209.87 $ & $68$ \\
 \hline
BM10 & " & " & $485$ & $500$ & $2$ & $0.00$ & $10^5$ & $0.258$ & $.258$ & $-0.23$ & $0.49$ & $0$ & $153.27 $ & $53$ \\
 \hline
BM11 & " & " & $485$ & $500$ & $2$ & $0.07$ & $10^5$ & $1.28$ & $0.002$ & $0.21$ & $0.49$ & $0$ & $169.28 $  & $60$ \\
 \hline
BM12 & " & " & $485$ & $500$ & $10$ & $0.0$ & $24752.5$ & $0.258$ & $0.258$ & $-0.23$ & $0.49$ & $0$ & $160.51 $  & $56$ \\
 \hline
 BM13 & " & " & $485$ & $500$ & $10$ & $0.1$ & $24752.5$ & $4.13$ & $0.22$ & $3.66$ & $0.49$ & $0$ & $241.75 $  & $70$ \\
 \hline
BM14 & " & " & $485$ & $485$ & $2$ & $0$ & $10^5$ & $0.258$ & $0.258$ & $-0.23$ & $0.24$ & $0.24$ & ${ 159.76 }$  & $59$ \\
 \hline
BM15 & " & " & $485$ & $485$ & $2$ & $0.07$& $10^5$ & $1.28$ & $0.002$ & $0.21$ & $0.244$ & $0.244$  & ${ 168.61}$ & $59$ \\
 \hline
 BM16 & " & " & $485$ & $485$ & $10$ & $0$ & $24752.5$ & $0.258$ & $0.258$ & $-0.23$ & $0.244$ & $0.244$ &${ 160.19 }$  & $56$\\
 \hline
BM17 &" & $485$ & $485$ & $485$ & $2$ & $0.0$ &$94090$ & $0.258$ & $0.258$ & $0.258$ & $0$ & $0$ &${ 161.31 }$  & $57$ \\
 \hline
BM18 & " & $485$ & $485$ & $485$ & $2$ & $0.07$ & $94090$ & $1.22$ & $0.02$ & $0.67$ & $0$ & $0$ & ${ 169.7 }$  & $60$ \\
 \hline
 BM19 & " &  $485$ & $485$ & $485$  & $10$ & $0$ & $23289.6$ & $0.258$ & $0.258$ & $0.258$ & $0$ & $0$ & ${ 160.96}$  & $57$  \\
 \hline
 BM20 & " &  $485$ & $485$ & $485$ & $10$ & $0.1$& $23289.6$ & $3.9$ & $0.22$ & $3.9$ & $0$ & $0$  & ${ 230.18 }$  & $70$ \\
 \hline
 BM21 & " &  $485$ & $485$ & $500$ & $2$ & $0$ & $94090$ & $0.258$ & $0.258$ & $0.258$ & $0$ & $0$ & ${ 161.31}$ & $57$  \\
 \hline
 BM22 & " &  $90$ & $200$ & $300$ & $2$ & $0$ & $3240$ & $0.258$ & $0.258$ & $1.31$ & $0.3$ & $-1.35$ & ${ 150.76 }$ & $51$  \\
 \hline 
  BM23 & " &  $90$ & $200$ & $300$ & $10$ & $0$ & $801.98$ & $0.258$ & $0.258$ & $1.31$ & $0.3$ & $-1.35$ & ${ 135.38  }$ & $37$  \\
 \hline 
  BM24 & " &  $90$ & $200$ & $300$ & $10$ & $0.2$ & $801.98$ & $0.263$ & $0.258$ & $1.06$ & $0.3$ & $-1.35$ & ${ 141.06  }$ & $42$  \\
 \hline 
		\end{tabular}}
	\end{center}
\end{table}

It is clear from figure \ref{f-entropy-1} that as the critical temperature increases, the entropy released into the primeval plasma also increases correspondingly. All these productions are higher than the entropy production in the SM, which is about $13\%$. The main contribution for 2HDM scenarios comes from the massive scalar bosons which are absent im the SM.

\section{Conclusion}
\label{sec.Conclusion}
We have shown, in this paper, that even in the minimal extension of SM with 2 Higgs bosons, EWPT is of first order and the net production of entropy is very large compared to the SM. An interesting point that is to be noted is $g_{\star}$ decreases as universe cools down. But as the temperature approaches to the decoupling temperatures of electrons, the contribution to the entropy becomes similar to that of SM.

Each and every benchmark points are calculated using BSMPT with the limiting condition $VEV/T_c>0.02$. This condition can be modified and it can be precised to even smaller one , which can give rise to more benchmark points. But overall, for the case of real, Type-I 2HDM, the effect will be similar and it will not vary. But this effect will change if other variants of 2HDM are considered.

This dilution will directly affect the pre-existing dark matter density of the universe, irrespective of the models in question. 
Not only this process, even phase transition at QCD epoch or the evaporation of PBHs produce entropy and can dilute pre-existing baryon asymmetry and dark matter density~\cite{Chaudhuri:2020wjo}. 
The extended standard model processes are under the observations at the present time.



\vspace{6pt}

\authorcontributions{Article by A.C., M.Y.K and S.P. AC and MYK contributed equally to this work. SP helped to prepare the manuscript.  The authors have read and agreed to the published version of the manuscript.
}

\acknowledgments{The work of AC and SP are funded by RSF Grant 19-42-02004.The research by M.K.was financially supported by Southern Federal University, 2020 Project VnGr/2020-03-IF.}

\conflictsofinterest{There has been no conflicts of interest among the authors.} 

\end{paracol}
\reftitle{References}



\begin{thebibliography}{99}







\bibitem{Abi:2021gix}
%
Abi, B., Albahri, T., Al-Kilani, S., Allspach, D., Alonzi, L.P., Anastasi, A., and, ... 
Measurement of the Positive Muon Anomalous Magnetic Moment to 0.46~ppm.
{\em \prl} {\bf  2021},
126, 141801. 
[\href{https://doi.org/10.1103/PhysRevLett.126.141801}{CrossRef}]
[\href{https://inspirehep.net/literature/1856627}{\Porey{ Inspire HEP}}]







\bibitem{Arcadi:2021yyr}
Arcadi, G., de Jesus, {\'A}.S., de Melo, T.B., Queiroz, F.S., and Villamizar, Y.S. 
A 2HDM for the g-2 and Dark Matter.
{\em arXiv e-prints} {\bf 2021}, arXiv:2104.04456.
[\href{https://inspirehep.net/literature/1857561}{\Porey{ Inspire HEP}}]









\bibitem{Chen:2021jok}
%
Chen, C.-H., Chiang, C.-W., and Nomura, T. 
Muon $g-2$ in two-Higgs-doublet model with type-II seesaw mechanism
{\em arXiv e-prints} {\bf 2021}, arXiv:2104.03275.
[\href{https://inspirehep.net/literature/1856548}{\Porey{ Inspire HEP}}]



\bibitem{Tanabashi:2018oca}
[Particle Data Group] Tanabashi, M., Hagiwara, K., Hikasa, K., Nakamura, K., Sumino, Y., Takahashi, F., and, ... 
Review of Particle Physics,
{\em \prd} {\bf 2018}, 98, 030001. 
[\href{https://doi.org/10.1103/PhysRevD.98.030001}{CrossRef}]
[\href{https://inspirehep.net/literature/1688995}{\Porey{ Inspire HEP}}]






\bibitem{Lunardini:2019zob}
Lunardini, C. and Perez-Gonzalez, Y.F. 
Dirac and Majorana neutrino signatures of primordial black holes.
{\em \jcap} {\bf 2020}, 014. 
[\href{https://doi.org/10.1088/1475-7516/2020/08/014}{CrossRef}]
[\href{https://inspirehep.net/literature/1759568}{\Porey{ Inspire HEP}}]








\bibitem{Endo:2021zal}
%
Endo, M., Hamaguchi, K., Iwamoto, S., and Kitahara, T. 
Supersymmetric Interpretation of the Muon $g-2$ Anomaly.
{\em arXiv e-prints} {\bf 2021}, arXiv:2104.03217.
[\href{https://inspirehep.net/literature/1856542}{\Porey{ Inspire HEP}}]







\bibitem{Katz:2014bha}
%
Katz, A. and Perelstein, M. 
Higgs Couplings and Electroweak Phase Transition. 
{\em \jhep
} {\bf 2014}, 108. 
[\href{https://doi.org/10.1007/JHEP07(2014)108}{CrossRef}]
[\href{https://inspirehep.net/literature/1276471}{\Porey{ Inspire HEP}}]



\bibitem{AS}
Sakharov, A.D. Violation of CP Invariance, C Asymmetry, and Baryon Asymmetry of the Universe. \emph{J. Exp. Theor. Phys. Lett.}  \textbf{1967}, \emph{5}, 24--27.









\bibitem{Karmakar:2020mds}
Karmakar S.
Relaxed Constraints on Masses of New Scalars in 2HDM.
Springer Proc. Phys. {\bf 2020} 248, 193-198 
[\href{https://doi.org/10.1007/978-981-15-6292-1\_23}{CrossRef}]
[\href{https://inspirehep.net/literature/1816177}{\Porey{ Inspire HEP}}]









\bibitem{Eberhardt:2020dat}
Eberhardt, O., Mart{\'\i}nez, A.P., and Pich, A. 
Global fits in the Aligned Two-Higgs-Doublet model.
{\em \jhep} {\bf 2021}, 5. 
 [\href{https://doi.org/10.1007/JHEP05(2021)005}{CrossRef}]
[\href{https://inspirehep.net/literature/1837110}{\Porey{ Inspire HEP}}]



\bibitem{Chaudhuri:2021agl}
Chaudhuri, A. and Khlopov, M.Y. 
Entropy production due to electroweak phase transition in the framework of two Higgs doublet model. 
{\em Physics} {\bf 2021}, 3, 275. 
[\href{https://doi.org/10.3390/physics3020020}{CrossRef}]
[\href{https://inspirehep.net/literature/1850436}{\Porey{ Inspire HEP}}]







\bibitem{Barroso:2013zxa}
Barroso, A., Ferreira, P.M., Santos, R., Sher, M., and Silva, J.P. 
2HDM at the LHC - the story so far.
{\em arXiv e-prints},
{\bf 2013},
arXiv:1304.5225.
[\href{https://inspirehep.net/literature/1228915}{\Porey{ Inspire HEP}}]




\bibitem{Karmakar}
Karmakar, S., Rakshit, S. 
Effective route beyond the extended scalar sectors of the standard model (Doctoral dissertation, Discipline of Physics, IIT Indore).
{\bf 2020}
[\href{http://dspace.iiti.ac.in:8080/jspui/handle/123456789/2304}{CrossRef}]



\bibitem{CarcamoHernandez:2021iat}
C{\'a}rcamo Hern{\'a}ndez, A.E., Espinoza, C., G{\'o}mez-Izquierdo, J.C., and Mondrag{\'o}n, M. 
Fermion masses and mixings, dark matter, leptogenesis and $g-2$ muon anomaly in an extended 2HDM with inverse seesaw.
{\em arXiv e-prints} {\bf 2021}, arXiv:2104.02730.
[\href{https://inspirehep.net/literature/1856562}{\Porey{ Inspire HEP}}]






\bibitem{Boeckel:2007tz}
Boeckel, T. and Schaffner-Bielich, J. 
Cosmology of fermionic dark matter.
{\em \prd} {\bf 2007}, 76, 103509. 
[\href{https://doi.org/10.1103/PhysRevD.76.103509}{CrossRef}]
[\href{https://inspirehep.net/literature/756435}{\Porey{Inspire HEP}}]







\bibitem{Chaudhuri:2017icn}
Chaudhuri, A. and Dolgov, A. 
Electroweak phase transition and entropy release in the early universe.
{\em \jcap} {\bf 2018}, 032. 
[\href{https://doi.org/10.1088/1475-7516/2018/01/032}{CrossRef}]
[\href{https://inspirehep.net/literature/1634633}{\Porey{ Inspire HEP}}]










\bibitem{Blinov:2015vma}
Blinov, N., Profumo, S., and Stefaniak, T.
The Electroweak Phase Transition in the Inert Doublet Model.
{\em JCAP
} {\bf 2015}, 028. 
[\href{https://doi.org/10.1088/1475-7516/2015/07/028}{CrossRef}]
[\href{https://inspirehep.net/literature/1364501}{\Porey{ Inspire HEP}}]







\bibitem{Chaudhuri:2020wjo}
Chaudhuri, A. and Dolgov, A.\ PBH evaporation, baryon asymmetry,and dark matter,
{\em arXiv e-prints} {\bf 2020}, arXiv:2001.11219.
[\href{https://inspirehep.net/literature/1777958}{\Porey{ Inspire HEP}}]


\end{thebibliography}
\end{document}